\documentclass[prd,twocolumn,showpacs,preprintnumbers,superscriptaddress,amsmath,amssymb,floatfix,nofootinbib]{revtex4-1}

\usepackage{graphicx,,booktabs}

\usepackage{amssymb}
\usepackage{amsmath,txfonts}
\usepackage{graphicx}
\usepackage{dcolumn}
\usepackage{color}
\usepackage{bm}
\usepackage[colorlinks,
            citecolor=blue,
            anchorcolor=green,
            menucolor=orange,
            linkcolor=red,
            filecolor=red,
            runcolor=pink,
            urlcolor=blue,
            frenchlinks=red]{hyperref}

\begin{document}

\title{\boldmath Productions of $f_{1}(1420)$ in pion and kaon  induced
reactions}
\author{Xiao-Yun Wang}
\thanks{xywang01@outlook.com}
\affiliation{Department of physics, Lanzhou University of Technology,
Lanzhou 730050, China}
\author{Jun He}
\thanks{Corresponding author : junhe@njnu.edu.cn}
\affiliation{Department of  Physics and Institute of Theoretical Physics, Nanjing Normal University,
Nanjing, Jiangsu 210097, China}
\author{Quanjin Wang}
\affiliation{Lanzhou University of Technology,
Lanzhou 730050, China}
\author{Hao Xu}
\affiliation{Department of Applied Physics, School of Science, Northwestern Polytechnical University, Xi'an 710129, China}
\begin{abstract}
The $f_{1}(1420)$ productions via pion and kaon induced reactions on a proton
target are investigated in an effective Lagrangian approach. Two treatments
of the $t$-channel Born term, the Feynman model and the Regge model,
are introduced to calculate the total and differential cross sections of the $\pi
^{-}p\rightarrow f_{1}(1420)n$ and $K^{-}p\rightarrow f_{1}(1420)\Lambda $ reactions.
The numerical results indicate that the experimental data of the total cross
section of the $\pi ^{-}p\rightarrow f_{1}(1420)n$ reaction can be reproduced
in both the Feynman and the Regge models. Based on the
parameters determined in the pion induced reaction, the cross sections of the $K^{-}p\rightarrow
f_{1}(1420)\Lambda $ reaction, about which there is little data found in the literature, are
predicted in the same beam momentum range. It is found that the line shapes of the total
cross section obtained in a kaon induced reaction with two treatments are
quite different. The cross sections for both reactions are at an order of
magnitude of $\mu$b, or larger, at a beam momenta up to 10 GeV/c. The differential
cross sections for both pion and kaon induced reactions are also present. It
is found that in the Regge model, the $t$ channel provides a
sharp increase at extreme forward angles. The results suggest that the
experimental study of the $f_1(1420)$ in the kaon induced reaction on a proton
target is as promising as in the pion induced reaction. Such an experimental
measurement is also very helpful to clarify the production mechanism of the $%
f_1(1420)$.
\end{abstract}

\maketitle

\section{Introduction}

The study of light mesons is an important way to understand the nonperturbative QCD.
Many light mesons have been observed and listed in the Review of Particle
Physics (PDG)~\cite{Tanabashi:2018oca}. However, the internal structure of
light mesons is still a confusing problem due to large nonperturbative effects
in the light flavor sector. Currently, the electron-positron collision is the
most important way to study  light mesons. It will be very helpful to
study the production and property of light mesons in different reactions.
A new detector, glueX, was equipped at  CEBAF after a 12 GeV upgrade,
which will focus on light meson spectroscopy~\cite{Austregesilo:2018mno}. Pion-induced light meson production is very important in the history of the
discovery of many light mesons. The secondary pion beam is accessible at
J-PARC~\cite{Kumano:2015gna} and COMPASS~\cite{Nerling:2012er} with
high intensity. The kaon beam can be also used to study light mesons, and it
is available at OKA@U-70~\cite{Obraztsov:2016lhp}, SPS@CERN~\cite%
{Velghe:2016jjw}, and J-PARC~\cite{Nagae:2008zz}. The data from future experiments at those facilities
will provide a good opportunity to deepen our understanding of the internal structure
of light mesons.

In the PDG, the $f_{1}(1420)$ is listed as an axial-vector state with
quantum numbers $I^{G}(J^{PC})=0^{+}(1^{++})$ with a suggested mass of 1426.4$%
\pm $ 0.9~MeV and a suggested width of 54.9$\pm $2.6~MeV~\cite%
{Tanabashi:2018oca}. The $f_{1}(1420)$ meson was first observed in
a pion nucleon interaction  in the Lawrence Radiation Laboratory in 1967~%
\cite{Dahl:1967pg}, and it was confirmed in other experiments with pion beams around
the year 1980~\cite{Corden:1978cz,Dionisi:1980hi,Bityukov:1983cw}. The $%
f_{1}(1420)$ was also observed in  recent experiments in $e^{+}e^{-}$ and
$J/\psi $ decays~\cite{Abdallah:2003gu,Achard:2007hm,Bai:1990hs}. Though the $f_{1}(1420)$ is well
established experimentally as a resonance structure, the
internal structure of the $f_{1}(1420)$ was, untill now, still unclear. In the
conventional $q\bar{q}$ picture, the $f_{1}(1420)$ can be classified as a
partner to the $f_{1}(1285)$ in the $^{3}P_{1}$ nonet of axial mesons, and
the mixture of nonstrange $f_{1q}=(u\bar{u}+d\bar{d})/\sqrt{2}$ and hidden-strange $%
f_{1s}=s\bar{s}$ was also discussed in the literature~\cite%
{Liu:2016rqu,Close:1997nm,Li:2000dy}.  However, a
recent study in Ref.~\cite{Debastiani:2016xgg} suggested that the $%
f_{1}(1420)$ is not a genuine resonance but results from the decay modes of the $%
f_{1}(1285)$ in $K^{\ast }\bar{K}$ and $\pi a_{0}(980)$ channels.

To determine the origin of the $f_1(1420)$, more precise experimental data are
required. In this work, based on the existent old data, we will analyze the $%
f_{1}(1420)$ production in pion induced reactions in an effective Lagrangian
approach. The kaon induced production will be discussed based on the results
of the pion induced interaction, which will be helpful for future high-precision  experimental
studies. Since in the current work we focus on the production mechanism of the $%
f_1(1420)$, the coupling constants are still determined with an assumption that  the $%
f_1(1420)$ is a genuine resonance~\cite{Tanabashi:2018oca}.

This paper is organized as follows: After the introduction, we present the 
formalism including Lagrangians and amplitudes of the $f_{1}(1420)$
productions in Sec. II. The numerical results of the cross section follow in
Sec. III. Finally, the paper{\ ends} with a brief summary.

\section{Formalism}

\begin{figure}[tbph]
\begin{center}
\includegraphics[scale=0.55]{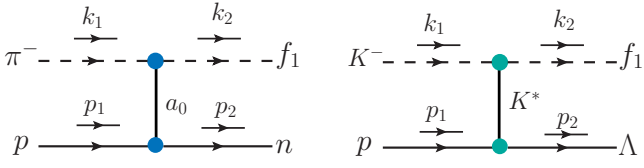}
\end{center}
\caption{Feynman diagrams for the $\protect\pi%
^{-}p\rightarrow f_{1}(1420)n$ reaction (left) and for the $%
K^{-}p\rightarrow f_{1}(1420)\Lambda $  reaction (right).}
\label{Fig: Feynman}
\end{figure}

The reaction mechanisms of pion and kaon induced productions of the $f_1(1420)$ are illustrated in Fig.~\ref{Fig: Feynman}. Usually,
the contribution from the $s$ channel with a nucleon pole is expected to be very
small, and it will be neglected in the current calculation. The $u$-channel
contribution is usually small and negligible at low energies~\cite%
{Wang:2017qcw}. Considering that the experimental data points at high energies
were obtained by the continuation of the $t$-channel contribution at very forward
angles to all angles, it is reasonable to calculate the cross section 
with only the $t$-channel contribution. In the present work, we do not
include the contributions from the nucleon resonances in the $s$ and $u$
channels.

Since the dominant decay of the $f(1420)$ is $K\bar{K}^*$, it is reasonable to
take  $K^*$ exchange as the dominant contribution in the $t$ channel of
kaon induced production. For pion induced production, we need the vertex for
decay of the $f_1(1420)$ with a pion. In Ref.~\cite{Barberis:1998by}, a
branch ratio of about 5\% was reported in the $a_0\pi$ channel. Hence, in the
current work, we adopt $a_0$ exchange in the $t$ channel of pion induced
production as in the case of pion induced $f_1(1285)$ production~\cite%
{Wang:2017qcw}.

\subsection{Lagrangians}

For pion induced production of the $f_{1}(1420)$ to describe the $t$%
-channel $a_{0}(980)$ ($\equiv a_{0}$) exchange we need the  following
Lagrangians \cite{Kochelev:2009xz,Domokos:2009cq,
Liu:2008qx,Colangelo:2010te},
\begin{eqnarray}
\mathcal{L}_{a_{0}NN} &=&g_{a_{0}NN}\bar{N}(\bm{\tau}\cdot \bm{a}_{0}){N},
\\
\mathcal{L}_{f_{1}a_{0}\pi } &=&-g_{f_{1}a_{0}\pi }f_{1}^{\mu} \partial
_{\mu }\bm{\pi}\cdot\bm{a}_{0} ,
\end{eqnarray}%
where ${N}$, ${f_{1}}$, $a_{0}$ and $\pi $ are the nucleon, $f_{1}(1420)$, $%
a_{0}(980)$ and $\pi $ meson fields, respectively. As suggested in the PDG~\cite%
{Tanabashi:2018oca}, the average width of $f_{1}(1420)$ is $54.9$ MeV. In
addition, the branching fraction of $f_{1}(1420)$ decay to $a_{0}\pi $ was
determined to be $4\%$ in Ref.~\cite{Barberis:1998by}. Thus, one gets $%
\Gamma _{f_{1}\rightarrow a_{0}\pi }\simeq 2.2$ MeV, which leads to a value of coupling constant $%
g_{f_{1}a_{0}\pi }\simeq 2.72$. In order to reduce the number of the free
parameter, we take the best fitting value $g_{a_{0}NN}=28.44$ in the
pion induced $f_{1}(1285)$ production~\cite{Wang:2017qcw}.

For kaon induced production, the relevant Lagrangians for the  $t$ channel
read \cite{Colangelo:2010te,Wan:2015gsl,Wang:2015xwa},
\begin{eqnarray}
\mathcal{L}_{f_{1}K^{\ast }K} &=&\frac{g_{f_{1}K^{\ast }K}}{m_{f_{1}}}%
(\partial _{\mu }K_{\nu }^{\ast }\partial ^{\mu }Kf_{1}^{\nu }-\partial
_{\mu }K_{\nu }^{\ast }\partial ^{\nu }Kf_{1}^{\mu }), \\
\mathcal{L}_{K^{\ast }N\Lambda } &=&-g_{K^{\ast }N\Lambda }\bar{N}\left( %
\rlap{$\slash$}K^{\ast }-\frac{\kappa _{K^{\ast }N\Lambda }}{2m_{N}}\sigma
_{\mu \nu }\partial ^{\nu }K^{\ast \mu }\right) \Lambda +\text{h.c.}~,
\end{eqnarray}%
where $m_{f_{1}}$ is the mass of $f_{1}$ meson,  and $K$, $K^{\ast }$, $f_{1}$, $N
$, and $\Lambda $ are the kaon, $K^{\ast }$, $f_{1}(1420)$, $\Lambda $, and the
nucleon fields, respectively. Here, we adopt the coupling constants $g_{K^{\ast
}N\Lambda }=-4.26$ and $\kappa _{K^{\ast }N\Lambda }=2.66$ as suggested by the
Nijmegen potential~\cite{Stoks:1999bz}.

The value of $g_{f_{1}K^{\ast }K}$ can be determined from the decay width%
\begin{eqnarray}
\Gamma _{f_{1}\rightarrow K^{\ast }K} &=&\left( \frac{g_{f_{1}K^{\ast }K}}{%
m_{f_{1}}}\right) ^{2}\frac{|\vec{p}_{K}^{~\mathrm{c.m.}}|}{24\pi
m_{f_{1}}^{2}}  \notag \\
&&\times \left[ \frac{(m_{f_{1}}^{2}-m_{K^{\ast }}^{2}-m_{K}^{2})^{2}}{2}%
+m_{K^{\ast }}^{2}E_{K}^{2}\right] ,
\end{eqnarray}%
with%
\begin{eqnarray}
|\vec{p}_{K}^{~\mathrm{c.m.}}| &=&\frac{\lambda(m_{f_{1}}^{2},m_{K^{\ast }}^{2},m_{K}^{2})}{2m_{f_{1}}}, \\
E_{K} &=&\sqrt{|\vec{p}_{K}^{~\mathrm{c.m.}}|^{2}+m_{K}^{2}},
\end{eqnarray}%
where $\lambda $ is the K$\ddot{a}$llen function with a definition of $\lambda
(x,y,z)=\sqrt{(x-y-z)^{2}-4yz}$. Since the branching fraction of $f_{1}(1420)$
decay to $K^{\ast }K$ was suggested to be $96\%$ in Ref.~\cite%
{Barberis:1998by}, one gets $g_{f_{1}K^{\ast }K}/m_{f_{1}}\simeq 8.36$ by
taking $\Gamma _{f_{1}\rightarrow K^{\ast }K}\simeq 52.7$ MeV \cite%
{Tanabashi:2018oca,Barberis:1998by}.

For the $t$-channel exchange \cite{Liu:2008qx}, the form factor $%
F(q^2)=(\Lambda ^{2}-m^{2})/(\Lambda^{2}-q^{2})$ is taken into account.
Here, $q$ and $m$ are the four-momentum and mass of the exchanged meson,
respectively. The value of the cutoff $\Lambda$ will be{\ determined by fitting the
experimental data}.

\subsection{Amplitudes}

According to the above Lagrangians, the scattering amplitude of the $\pi
^{-}p\rightarrow f_{1}(1420)n$ or $K^{-}p\rightarrow f_{1}(1420)\Lambda $
reaction can be written as%
\begin{equation}
-i\mathcal{M}=\epsilon _{f_{1}}^{\mu \ast }(k_{2})\bar{u}(p_{2})\mathcal{%
A}_{\mu }u(p_{1}),
\end{equation}%
where $\epsilon _{f_{1}}^{\mu \ast }$ is the polarization vector of $f_{1}$
meson, and $\bar{u}$ or $u$ is the Dirac spinor of nucleon or $\Lambda $
baryon.

For the $\pi ^{-}p\rightarrow f_{1}(1420)n$ reaction, the reduced amplitude
$\mathcal{A}_{i,\mu }$  reads
\begin{equation}
\mathcal{A}_{\mu }^{(a_{0})}=i\sqrt{2}g_{a_{0}NN}g_{f_{1}a_{0}\pi }F(q^2)%
\frac{1}{t-m_{a_{0}}^{2}}k_{1\mu },  \label{AmpT1}
\end{equation}%
where $t=(k_{1}-k_{2})^{2}$ is the Mandelstam variables. The coupling
constants are fixed with the experimental data and the fitting of the pion induced $f_1(1285)$ production as addressed above.

For the  $K^{-}p\rightarrow f_{1}(1420)\Lambda $ reaction, the reduced
amplitude $\mathcal{A}_{\mu }$ is written as
\begin{eqnarray}
\mathcal{A}_{\mu }^{(K^{\ast })} &=&ig_{K^{\ast }N\Lambda }\frac{%
g_{f_{1}K^{\ast }K}}{m_{f_{1}}}F(q^2)\left( \gamma _{\xi }-i\frac{\kappa
_{K^{\ast }N\Lambda }}{2m_{N}}\gamma _{\xi }\rlap{$\slash$}q_{K^{\ast
}}\right)  \notag \\
&&\frac{\mathcal{P}^{\nu \xi }}{t-m_{K^{\ast }}^{2}}\left[
(k_{1}-k_{2})\cdot k_{1}g_{\mu \nu }-(k_{1}-k_{2})_{\mu }\cdot k_{1\nu }%
\right] ,  \label{AmpT2}
\end{eqnarray}%
with
\begin{equation}
\mathcal{P}^{\nu \xi }=i\left( g^{\nu \xi }+q_{K^{\ast }}^{\nu }q_{K^{\ast
}}^{\xi }/m_{K^{\ast }}^{2}\right).
\end{equation}
Here, the coupling constants are also fixed as in the pion induced
production. Hence, the only free parameter is the cutoff in form factor.

\subsection{Reggeized $t$-channel}\label{Sec: Regge}

To analyze hadron production at high energies, a more economical approach
may be provided by a Reggeized treatment \cite%
{Wan:2015gsl,Wang:2015xwa,Haberzettl:2015exa,Wang:2015hfm,Galata:2011bi,Ozaki:2009wp}%
. In our previous works \cite{Wang:2015hfm,Wang:2017qcw,Wang:2017plf}, an
interpolating Reggeized treatment was introduced to interpolate the Regge
trajectories smoothly to the Feynman propagator at low energies, as proposed in Ref.~\cite{Nam:2010au} . Because there
are only 4 data points, we do not adopt the interpolating Reggeized
treatment, but we discuss both the Feynman model and the Regge model. In the
Feynman model, the $t$-channel amplitude in Eqs.~(\ref{AmpT1}) and (\ref%
{AmpT2}) is applied directly. The Regge model can be introduced by replacing
the $t$-channel Feynman propagator with the Regge propagator as,
\begin{eqnarray}
\frac{1}{t-m_{a_{0}}^{2}} &\rightarrow &(\frac{s}{s_{scale}})^{\alpha
_{a_{0}}(t)}\frac{\pi \alpha _{a_{0}}^{\prime }}{\Gamma \lbrack 1+\alpha
_{a_{0}}(t)]\sin [\pi \alpha _{a_{0}}(t)]}, \\
\frac{1}{t-m_{K^{\ast }}^{2}} &\rightarrow &(\frac{s}{s_{scale}})^{\alpha
_{K^{\ast }}(t)-1}\frac{\pi \alpha _{K^{\ast }}^{\prime }}{\Gamma \lbrack
\alpha _{K^{\ast }}(t)]\sin [\pi \alpha _{K^{\ast }}(t)]}.
\end{eqnarray}%
The scale factor $s_{scale}$ is fixed at 1 GeV$^2$. In addition, the Regge
trajectories of $\alpha _{a_{0}}(t)$ and $\alpha _{K^{\ast }}(t)$ read as \cite%
{Galata:2011bi,Ozaki:2009wp},%
\begin{equation}
\alpha _{a_{0}}(t)=-0.5+0.6t/{\rm GeV}^2,\ \alpha _{K^{\ast }}(t)=1+0.85(t-m_{K^{\ast
}}^{2})/{\rm GeV}^2.\quad \ \
\end{equation}
After the Reggeized treatment is introduced, no additional parameter is introduced.

\section{Numerical results}

With the preparation in the previous section, the cross section of the $\pi
^{-}p\rightarrow f_{1}(1420)n$ and $K^{-}p\rightarrow f_{1}(1420)\Lambda $
reactions will be calculated and compared with  the experimental data \cite%
{Dahl:1967pg,Corden:1978cz,Dionisi:1980hi,Bityukov:1983cw}. The differential
cross section in the center of mass (c.m.) frame is written as
\begin{equation}
\frac{d\sigma }{d\cos \theta }=\frac{1}{32\pi s}\frac{\left\vert \vec{k}%
_{2}^{{~\mathrm{c.m.}}}\right\vert }{\left\vert \vec{k}_{1}^{{~\mathrm{c.m.}}%
}\right\vert }\left( \frac{1}{2}\sum\limits_{\lambda }\left\vert \mathcal{M}%
\right\vert ^{2}\right) ,
\end{equation}%
where $s=(k_{1}+p_{1})^{2}$, and $\theta $ denotes the angle of the outgoing
$f_{1}(1420)$ meson relative to the $\pi $/$K$ beam direction in the c.m. frame.
$\vec{k}_{1}^{{~\mathrm{c.m.}}}$ and $\vec{k}_{2}^{{~\mathrm{c.m.}}}$ are
the three-momenta of the initial $\pi $/$K$ beam and final $f_{1}(1420)$,
respectively.

\subsection{Cross section of the $\protect\pi ^{-}p\rightarrow f_{1}(1420)n$
reaction}

In this work, we minimize $\chi ^{2}$ per degree of freedom ($d.o.f.$) for the experimental data of the total
cross section by fitting the cutoff parameter $%
\Lambda$ using a total of 4 data points at the beam momentum $p_{Lab}$ from
3.1 to 13.5 GeV/c. The fitted cutoff and the $\chi^2/dof$ are listed in Table~\ref{tab:fit1}.
\renewcommand\tabcolsep{0.82cm} \renewcommand{\arraystretch}{1.2}
\begin{table}[h]
\caption{The fitted value of free parameter $\Lambda _{t}$ in the unit of
GeV.}
\label{tab:fit1}%
\begin{tabular}{lcc}
 \toprule[1.5pt]
& $\Lambda$ & $\chi ^{2}/dof$ \\ \hline
Feynman & $1.18\pm 0.01$ & $1.68$ \\
Regge & $1.60\pm 0.03$ & $1.93$ \\  \bottomrule[1.5pt]
\end{tabular}%
\end{table}

\begin{figure}[tbph]
\begin{center}
\includegraphics[bb=50 10 680 540,scale=0.34,clip]{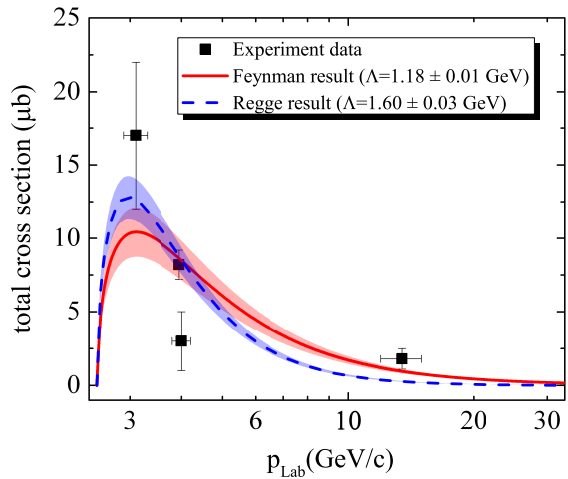}
\end{center}
\caption{Total cross section for the $\protect\pi %
^{-}p\rightarrow f_{1}(1420)n$ reaction. The Full (red) and dashed (blue)
lines are for the Feynman  and Regge models, respectively. The bands
stand for the error bar of the cutoff $\Lambda.$  The experimental data are from Refs.~\cite{Dahl:1967pg,Corden:1978cz,Dionisi:1980hi,Bityukov:1983cw}.}
\label{Fig:total01}
\end{figure}

  From Fig.~\ref%
{Fig:total01}, it is found that the experimental data of the total cross section for the $\pi
^{-}p\rightarrow f_{1}(1420)n$  reaction is reproduced in both the Feynman
 and the Regge models. The line shapes of total cross section in  both models are analogous.  The total cross section increases sharply near the threshold  and reaches a maximum at a beam momentum of about 3 GeV/c. The Regge model gives a  cross section that is a little larger  than the Feynman model  in this momentum range.  The total cross section decreases at momenta larger than 3 GeV/c in both the Feynman and the Regge models. In this momentum range, the Feynman model gives a larger total cross section. The $\chi^2/dof$ are 1.68 and 1.93 for the Feynman  and  the Regge models, respectively.   Both models can describe the existent rough data quantitatively. Based on the current results, it can be expected that the data will be described better if  a mixing of two models,  that is, the
interpolating Regge treatment \cite{Wang:2015hfm,Wang:2017qcw,Wang:2017plf, Nam:2010au},  is introduced, which requires more precise data.  From Fig.~\ref{Fig:total01}, the $\chi^2$ comes mainly from the two data
points around 4 GeV/c. In both models, the upper data point is roughly obtained,
which can be checked in future high-precision experiment.

\begin{figure}[h!]
\centering
\includegraphics[scale=0.38]{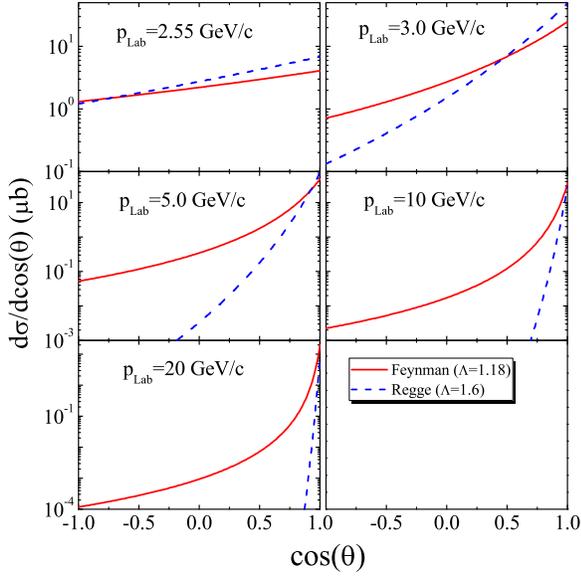}
\caption{The differential cross section $d\protect\sigma %
/d\cos \protect\theta $ of the $\protect\pi ^{-}p\rightarrow f_{1}(1420)n$
process as a function of $\cos \protect\theta $. The Full (red) and dashed
(blue) lines are for the results of Feynman model and Regge model,
respectively.}
\label{dcs01}
\end{figure}

In Fig.~\ref{dcs01}, we present our prediction of the differential cross
section of the $\pi ^{-}p\rightarrow f_{1}(1420)n$ reaction in two schemes at
different beam momenta. It can be seen that the differences between the
Regge  and the Feynman models at low momenta are small but become large at
higher momenta. With increase of the beam momentum, the slope of the curve in
the Regge model is steeper than that in the Feynman model at forward angles,
which can be tested by further experiment to clarify the role of the
Reggeized treatment.

\subsection{Cross section of the $K^{-}p\rightarrow f_{1}(1420)\Lambda $
reaction}

\begin{figure}[h!]
\begin{center}
\includegraphics[bb=50 10 680 540,clip,scale=0.34]{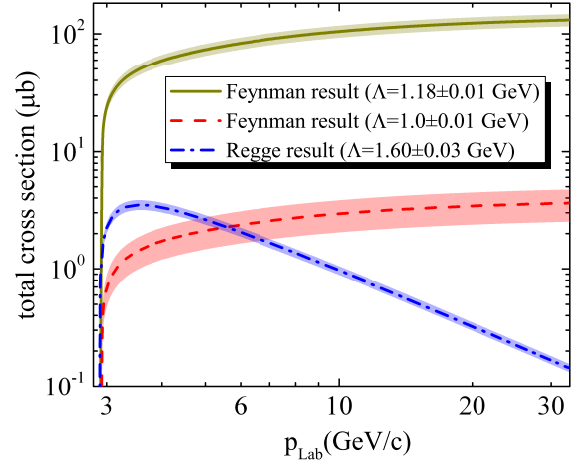}
\end{center}
\caption{Total cross section for $K^{-}p\rightarrow
f_{1}(1420)\Lambda $ reaction. The Full (red) and dashed (blue) lines are
for the Feynman model and Regge model, respectively. The bands stand for the
error bar of cutoff $\Lambda _{t}.$}
\label{Fig:total02}
\end{figure}

Since   there exists no  experimental data for the $K^{-}p\rightarrow
f_{1}(1420)\Lambda $ reaction,  here we  give the prediction of the
cross section for the $K^{-}p\rightarrow f_{1}(1420)\Lambda $ reaction. In
Ref.~\cite{Bityukov:1983cw}, the experiment shows that $\sigma
(K^{-}p\rightarrow f_{1}(1420)\Lambda )/\sigma (\pi ^{-}p\rightarrow
f_{1}(1420)n)>10$ at a beam momentum $p_{Lab}=32.5$ GeV/c.  Here, we make a comparison of this data with the prediction in the current work.  In our 
calculation above, the $\sigma (\pi ^{-}p\rightarrow f_{1}(1420)n)\simeq 0.017$ ($%
\mu $b) at $p_{Lab}=32.5$ GeV/c by taking a value of cutoff $\Lambda=1.6$ GeV in the Regge
model. For the $K^{-}p\rightarrow f_{1}(1420)\Lambda $ reaction, by taking a value of the cutoff $%
\Lambda=1.6$ GeV, one can get a value of the total cross section about 0.14 $\mu $b
in the Regge model, which means that the value of the cutoff $\Lambda=1.6$ GeV is relatively
reasonable  to calculate the cross section of the $K^{-}p\rightarrow
f_{1}(1420)\Lambda $ reaction in the Regge model. In Fig.~\ref{Fig:total02}
we present the total cross section of the $K^{-}p\rightarrow f_{1}(1420)\Lambda $
reaction within the Regge model.

In the Feynman case, if we choose the value of the cutoff as in the pion induced case,
the cross section of the $K^{-}p\rightarrow f_{1}(1420)\Lambda $ reaction will
increase continuously with the increasing of the momentum $p_{Lab}$ in the
momentum range considered. Though it does not conflict with $\sigma
(K^{-}p\rightarrow f_{1}(1420)\Lambda )/\sigma (\pi ^{-}p\rightarrow
f_{1}(1420)n)>10$ as suggested in Ref.~\cite{Bityukov:1983cw}, it seems
unnatural that a total cross section will reach 100 $\mu $b.  Hence, here we chose a
smaller value of cutoff $\Lambda $, 1.0 GeV.

Different from the pion induced case, from Fig.~\ref{Fig:total02}, the line shape of total cross section in the
Regge model is quite different from that in the Feynman model. For the Regge
case, we notice that the line shape of the total cross section goes up very
rapidly and has a peak around $p_{Lab}=3.53$ GeV/c. For the Feynman case, the value of total cross section becomes larger and larger
with the increasing of the beam momentum up to 20 GeV/c. The monotonically
increasing behaviour may be caused by the $K^{\ast }$ exchange amplitude in the Feynman model~\cite%
{Ozaki:2009wp}.  Compared with the Regge model, as discussed in Sec.~\ref{Sec: Regge} the Feynman model  is less suitable to describe the behavior of the cross section at high beam momenta.  The differences between the Regge and the Feynman model will
be useful in clarifying the role of the Reggeized treatment in future experiment.
\begin{figure}[t]
\centering
\includegraphics[scale=0.38]{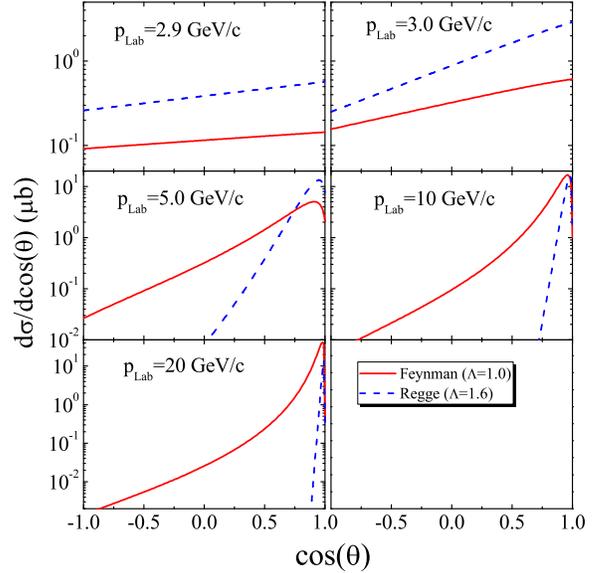}
\caption{The differential cross section $d\protect\sigma %
/d\cos \protect\theta $ of the $K^{-}p\rightarrow f_{1}(1420)\Lambda $
process as a function of $\cos \protect\theta $. The Full (red) and dashed
(blue) lines are for the results of Feynman model and Regge model,
respectively.}
\label{dcs02}
\end{figure}
\ \

The differential cross sections in the two models are illustrated in Fig.~\ref%
{dcs02}, which shows that the discrepancy of the differential cross sections
in the two models is small at low beam momenta but becomes large at higher beam
momenta. From Fig.~\ref{dcs02}, one notices that, comparing with the results
in the Feynman model, the differential cross section in the Regge model is very
sensitive to the $\theta $ angle and gives a considerable contribution at
forward angles with the increase of beam momentum.

\section{Summary and discussion}

We have studied the $\pi ^{-}p\rightarrow f_{1}(1420)n$ and $%
K^{-}p\rightarrow f_{1}(1420)\Lambda $ reactions within the Feynman  and
Regge models. For the $\pi ^{-}p\rightarrow f_{1}(1420)n$ reaction, both
results calculated in the  Regge and the Feynman models can be reproduced in the experimental
data, but the differential cross sections at high beam momenta are quiet different.
It is found that the differential cross sections for the $\pi ^{-}p\rightarrow
f_{1}(1420)n$ reaction in the Regge model are very sensitive to the $\theta $
angle and give  considerable contributions at forward angles, which can be
checked by future experiment and may be an effective way to examine the
validity of the Reggeized treatment.

For the $K^{-}p\rightarrow f_{1}(1420)\Lambda $ reaction, the line shapes of the total
cross sections obtained in both models are very different. The total cross
section increases continuously in the Feynman model in the beam momentum range considered
in the current work. If the cutoff of the  pion induced production is
adopted directly, the cross section increases to 100 $\mu$b. In the Regge model, it decreases at momenta larger than $p_{Lab}=4$ GeV/c. It is consistent with
 the Regge model being more suitable to describe the behavior of the cross
section at high momenta. The line shape of the differential cross section of the $K^{-}p\rightarrow f_{1}(1420)\Lambda $ reaction is similar to the result
of the $\pi ^{-}p\rightarrow f_{1}(1420)n$ reaction. In the Regge model, the $t$ channel provides a sharp increase at extreme forward
angles.

In the current work, the pion induced $f_1(1420)$ production is studied with the $a_0$ exchange (see Fig.~\ref{Fig: Feynman}).  The results suggest that the couplings of the $f_1(1420)$ with $a_0\pi$ should be considerably large.  To some extent, it supports the interpretation of the $f_1(1420)$ as a reflection of the decay modes of the $f_1(1285)$ in Ref.~\cite{Debastiani:2016xgg} where a large branch ratio, about 17\%, in the $a_0\pi$ channel is extracted. Of coarse, it does not mean that interpretations in the $s\bar{s}$ picture are disfavored, because there is no theoretical calculation about the $a_0\pi$ decay in the constituent quark model. To deepen the understanding of the $f_1(1420)$, experimental studies of the $f_1(1420)$ with the pion and kaon beams are strongly suggested. The results of this work suggest that it is promising to do such an experiment at existing facilities. The expected high-precision data will be very helpful to extract the couplings of the $f_1(1420)$ and $K^*\bar{K}$ and $a_0\pi$ with the method in this work. Besides, the subsequential decay that produced $f_1(1420)$ in such experiments will provide more information about the $f_1(1420)$.

The pion and kaon beams can be provided at J-PARC and COMPASS. The precision of  possible data in  future experiments at these facilities
will be much higher than in old experiments. The above theoretical results
may provide valuable information for possible experiments to study 
the $f_{1}(1420)$ at these facilities.

\section{Acknowledgments}

This project is supported by the National Natural Science Foundation of
China under Grants No. 11705076 and No. 11675228. We acknowledge the Natural
Science Foundation of Gansu province under Grant No. 17JR5RA113.

\end{document}